\documentclass[preprint,11pt,5p]{elsarticle}

\usepackage{amsmath}
\usepackage{amssymb}
\usepackage{bm}
\usepackage[retainorgcmds]{IEEEtrantools}
\usepackage{color}

\journal{Physics Letters B}

\begin{document}

\begin{frontmatter}



\title{Meson resonances in forward-angle $\pi^+\pi^-$ photoproduction}

\author{\L. Bibrzycki}
\address{Institute of Computer Science, Pedagogical University of Cracow, 
30-084 Krak\'ow, Poland}
\ead{lukaszb@up.krakow.pl}
\author{P. Byd\v{z}ovsk\'y}
\address{Nuclear Physics Institute, CAS, 25068 \v{R}e\v{z}, Czech
Republic}
\author{R. Kami\'nski}
\address{Institute of Nuclear Physics, Polish Academy of Sciences, Division of Theoretical Physics, 
31-342 Krak\'ow, Poland}
\author{A. P. Szczepaniak}
\address{Physics Department, Indiana University, Bloomington, IN 47405, USA}
\address{Center for Exploration of Energy and Matter, Indiana University, Bloomington, IN 47403, USA}
\address{Theory Center, Thomas Jefferson National Accelerator Facility,}

\begin{abstract}
Assuming that the $\pi^+\pi^-$ photoproduction at forward angles and high energies is dominated by 
 one pion exchange we calculate the $\pi^+\pi^-$ mass distributions for low partial waves.  
Predictions of the model agree well with 
the experimental data which indicate that the $S$, $P$ and $D$ waves are dominated by the 
 $f_0(980)$, $\rho(770)$ and $f_2(1270)$, resonances  respectively. 
\end{abstract}

\begin{keyword}
Photoproduction \sep Partial wave analysis \sep Final state interactions


\PACS  13.75.Lb \sep 13.60.Le\\
JLAB-THY-18-2808
\end{keyword}
\end{frontmatter}
%\textbf{\section{}
%\label{}
%
%%% The Appendices part is started with the command \appendix;
%%% appendix sections are then done as normal sections
%%% \appendix
%
%%% \section{}
%%% \label{}
%
%%% For citations use: 
%%%       \citet{<label>} ==> Jones et al. [21]
%%%       \citep{<label>} ==> [21]
%%%
%
%%% If you have bibdatabase file and want bibtex to generate the
%%% bibitems, please use
%%%
%%%  \bibliographystyle{elsarticle-num-names} 
%%%  \bibliography{<your bibdatabase>}
%
%%% else use the following coding to input the bibitems directly in the
%%% TeX file.
%
%\begin{thebibliography}{00}
%
%%% \bibitem[Author(year)]{label}
%%% Text of bibliographic item
%\end{thebibliography}
%\end{document}
%
% Introduction
%
%\section{Introduction}\label{introduction}
Photoproduction is an important reaction in hadron spectroscopy. To determine resonance production mechanisms one performs partial wave analysis of the differential cross section  in various final state channels. This is now possible thanks to availability of high-quality data from JLab, ELSA, MAMI, and SPring-8.  
Among those the CLAS data continues to be of high 
interest as it remains to be the only data on photoproduction of $f_0$ resonances.
 Specifically, from analysis of forward photoproduction of pseudoscalar mesons one can investigate the spectrum of light meson resonances, including those with exotic quantum numbers \cite{SwatSzczepaniak}, which are important for development of our understanding of color confinement.  In the previous studies  we have shown that  $S$ and $D$ resonances are copiously produced in di-pion  photoproduction~\cite{PRD-98,PRD-13}. 
%The proper description 
 %required inclusions of  
  %final-state interaction (FSI) between 
% mesons. 
In those studies we assumed that the di-pion photoproduction  
is dominated by the $t$-channel $\rho$ and $\omega$ exchanges at  the nucleon vertex. In the present work we focus instead on the general 
properties of the production process. Specifically 
 we examine two principal modes. The long-range mode related to the 
one pion exchange and the short-range one, which effectively takes into 
account all heavier meson exchanges and/or quark/gluon processes. 
As a function of the di-pion mass, the latter has singularities far away 
from the physical region and can be parametrized it terms of a suitably chosen smooth functions. These two modes naturally arise when one considers restrictions imposed by unitarity on  final state interactions in a general production process~\cite{Aitchison}.
 Instead of assuming a particular exchange mechanism, 
 we generalize the conventional formulation of the Deck model \cite{Deck,Pumplin} by applying the phenomenological set of pion-nucleon  amplitudes obtained by the SAID group \cite{SAID} and to describe the final state interactions in the $\pi\pi$ channel we use a set of partial wave  amplitudes from a recent analysis in ~\cite{BKN2016}. The use of phenomenological  $\pi N$ and $\pi\pi$ amplitudes  enables us to make
  a prediction for the absolute normalization of the long range mode of the  photoproduction amplitude, while the short range mode is fitted to the data. Resulting cross sections, as we show in this paper, agree well with the available data on the $\pi^+\pi^-$  photoproduction in the $S$, $P$, $D$, and $F$ waves.

\begin{figure}[htb]
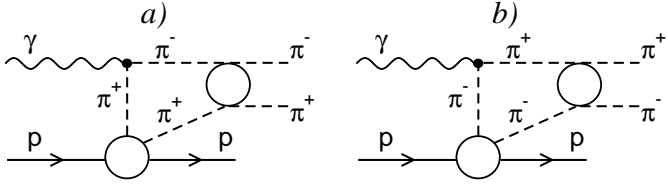

\begin{center}
\includegraphics[angle=0,scale=0.17]{twopion_SAID-a.eps}\hspace{4mm}
\includegraphics[angle=0,scale=0.17]{twopion_SAID-b.eps}
\end{center}
\caption{Diagrams for the pion photoproduction (Deck mechanism), where pions are subject to final state interactions.}
\label{BornScat}
\end{figure}

\begin{figure}[htb]
\begin{center}
\includegraphics[angle=270,scale=0.3]{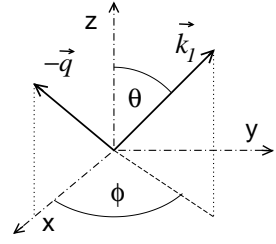}
\end{center}
\caption{Coordinate system in the $\pi\pi$ c.m. reference frame.}
\label{kinematics}
\end{figure}

{\it Model description.}  
For the $\pi^+\pi^-$ photoproduction on the proton
%\begin{equation}
$\gamma\,(q,\lambda) + p(p_1,\lambda_1) 
\rightarrow p(p_2,\lambda_2) + \pi^+\,(k_1) + \pi^-\,(k_2)$, where $\lambda$'s denote particle helicities, 
%\label{process}
%\end{equation}
the invariant amplitude is related to the S matrix by 
\begin{equation}
S_{fi}=\delta_{fi}+i(2\pi)^4\,\delta^4(p_2 +k_1+k_2-p_1-q)\,{\cal T}_{fi}\,.
\label{smatrix5}
\end{equation}
Accordingly, the invariant double-differential cross section expressed as a sum over $\pi\pi$ partial waves is given by 
%(in the normalization $\bar{u}5(p_2,\lambda_2)\,u(p_1,\lambda_1)=2m$) is
\begin{equation}
\frac{d^2\sigma}{d|t|\,d\sqrt{s_{\pi\pi}}} = \frac{1}{64(2\pi)^4}\frac{|{\bf k}|}{(s-m^2)^2 }\,
\sum_{lm}
\sum_{\lambda_2\lambda\lambda_1} 
|{\cal T}^{lm}|^2\,,%|{\cal T}^{lm}|^2\,,
\label{crs}
\end{equation} 
%where $p$ and $q$ are definded in the Lab frame %($\bm{p}\!=\!0$), $q\cdot p = %m\,|\bm{q}^{\,Lab}|$.
where $|{\bf k}| = \sqrt{s_{\pi\pi}/4-m_\pi^2}$
 is the magnitude of pion momenta in the $\pi\pi$ rest frame. 
The partial wave projection is defined 
 in the $\pi\pi$ center of mass frame { \it cf.} 
  Fig.~\ref{BornScat}. 
In this frame the direction of the recoil proton defines the negative $z$ axis and $y$ axis is perpendicular the di-pion production plane.
The orientation of the $\pi^+$ momentum is given by the polar and azimuthal angles, 
 $\theta$ and $\phi$ as shown in Fig.~\ref{kinematics}, with the photon momentum 
  given by $\bm{q}=|\bm{q}|(-\sin\theta_q,0,\cos\theta_q)$ 
 where $|\bm{q}|=(s_{\pi\pi}-t)/2 \sqrt{s_{\pi\pi}}$ and $\cos\theta_q$ is an algebraic function of the Mandelstam invariants. 

In terms of the scattering amplitude ${\cal T}$ 
 the partial wave amplitudes are given by 
%\begin{multline}
\begin{equation}
{\cal T}^{lm} = 
\int d\Omega\;Y^*_{lm}(\Omega)\;
{\cal T} (p_2 \lambda_2\,k_1 k_2, q \lambda\,p_1 \lambda_1)
\end{equation} 
where $d\Omega = d\,{\cos}\theta \,d\phi$. The partial wave amplitudes  depend on the total invariant energy 
$s=(q+p_1)^2$, momentum transfer $t=(p_2-p_1)^2$, and $\pi\pi$ invariant mass $\sqrt{s_{\pi\pi}}$. 
A similar expression holds for the Deck amplitude ${\cal M}^{lm}$ (see below).

For each spin, $l$ and isospin, $I=0,1,2$ the final state interactions are described by the $\pi\pi$ partial wave amplitudes, $t^I_l$ that are given by the phase shifts $\delta_l^I$ and inelasticity parameters  $\eta_l^I$,
\begin{equation}
t_l^I=\frac{1}{2i\rho}\left(\eta_l^I\,{\rm e}^{2i\delta_l^I} - 1\right),  
\label{eq_on-shell}
\end{equation}
where $\rho = 2|{\bf k}|/\sqrt{s_{\pi\pi}}$. 
The partial wave amplitudes $t_l^I(s_{\pi\pi})$ are taken from the recent study of \cite{BKN2016}, where crossing symmetry and once subtracted dispersion relations were imposed to further constrain 
 the amplitudes that were studied previously in ~\cite{BKN2014,YuSu2010,GKPYIV}.

 In the limit of a large production range, the partial waves are related to the FSI amplitudes by a simple algebraic relation~\cite{Aitchison}, which  for the even waves, assuming isospin symmetry reads, 
  \begin{multline}
{\cal T}_{\pi^+\pi^-}^{lm}(\lambda_2\,\lambda\,\lambda_1) = \\
\left[ 1 + i\rho\,\left(\frac{2}{3} t_l^{0} +
\frac{1}{3} t_l^{2}\right)\right] 
{\cal M}_{\pi^+\pi^-}^{lm}(\lambda_2\,\lambda\,\lambda_1) \,,
\label{eq_PRD-16}
\end{multline} 
and for the odd ones
\begin{equation}
{\cal T}_{\pi^+\pi^-}^{lm}
(\lambda_2\,\lambda\,\lambda_1) = 
\left[ 1 + i\rho\,t_l^{1}\right] 
{\cal M}_{\pi^+\pi^-}^{lm}
(\lambda_2\,\lambda\,\lambda_1).
\label{eq_PRD-162}
\end{equation} 
Here the long-range production,  ${\cal M}^{lm}$  
 is taken as the partial wave projection of one pion exchange {\it aka} the Deck amplitude.
%\label{secDeck}
The Deck amplitude was originally constructed in ~\cite{Deck} under the assumption that contribution from the nearest singularity at low-$t_{\gamma\pi}$, which is  the channel dual to 
$s_{\pi\pi}$, is that of the pion pole. Moreover, gauge invariance was imposed by modifying the pion pole according to a following prescription  ~\cite{Pumplin}, 
\begin{multline}
M_{\lambda_2\lambda\lambda_1}=-e\left[
\left(\frac{\epsilon_{\lambda}\cdot k_2}
{q\cdot k_2} -\frac{\epsilon_{\lambda}\cdot(p_1+p_2)}
{q\cdot (p_1+p_2)}\right)T^+_{\lambda_1\lambda_2}\right.\\ -\left. \left(
\frac{\epsilon_{\lambda}\cdot k_1}
{q\cdot k_1}-\frac{\epsilon_{\lambda}\cdot(p_1+p_2)}
{q\cdot (p_1+p_2)}\right)T^-_{\lambda_1\lambda_2}
\right]
\label{Deck}
\end{multline}
where $e$ is the electric charge, $\epsilon_\lambda$ is the photon helicity  polarization vector and 
$T^+_{\lambda_1\lambda_2}$ and $T^-_{\lambda_1\lambda_2}$ are $\pi^\pm N$ 
 scattering amplitudes.
 %Notation of the kinematics is given in 
 %Eq.~(\ref{process}). 
This is one of many possible implementations of gauge invariance. 
Another model, for example,  was studied in  \cite{Stichel} where contributions  from the baryon exchanges were also included, which 
 required a different modification  to make the overall amplitude gauge invariant. In the following we use Eq.~(\ref{Deck}), which appears better 
suited in the kinematics dominated by meson exchanges. 
 Similarly to ${\cal T}^{lm}$ the partial wave projection 
  of the Deck amplitude is given by, 
%  in the $\pi\pi$ c.m. reference frame, see Fig.~\ref{kinematics}, reads
\begin{equation} 
{\cal M}_{\pi^+\pi^-}^{lm}({\lambda_2\lambda\lambda_1})=\int \!d\Omega \,{Y^\ast_{lm}}(\Omega)\, M_{\lambda_2\lambda\lambda_1}. 
\label{pw-expansion}
\end{equation}
%where $M_{\lambda_2\lambda\lambda_1}$ is given in Eq.~(\ref{Deck}).
%
%\subsection{$\pi p$ elastic amplitudes}
Elastic amplitudes of the $\pi^+$ and $\pi^-$ scattering off protons that appear in Eg.~(\ref{Deck}) can be expressed in terms of the isospin amplitudes
\begin{equation}
T^+_{\lambda_1\lambda_2}=T^{\frac{3}{2}}_{\lambda_1\lambda_2},
\quad T^-_{\lambda_1\lambda_2}=\frac{1}{3}(T^{\frac{3}{2}}_{\lambda_1\lambda_2}+2 T^{\frac{1}{2}}_{\lambda_1\lambda_2}).
\label{charge-amp}
\end{equation}
with the latter given in terms of the standard Lorentz invariant 
isospin amplitudes~\cite{ChewGoldLowNamb}
\begin{equation}
T_{\lambda_1\lambda_2}^{I}=\overline{u}(p_2,\lambda_2)\left(A^{I}+\gamma\cdot QB^{I}\right)u(p_1,\lambda_1)
\end{equation}
with  $Q=\frac{1}{2}(q\mp k_1\pm k_2)$, for $\pi^-$ and $\pi^+$ scattering,  respectively. To construct the amplitudes
in Eq.~(\ref{charge-amp}) we use the SAID $\pi N$ partial wave parametrization. Note that due to kinematics of the process the pion that undergoes the scattering on the proton target is not on 
its mass shell: $(q-k_1)^2\neq m_{\pi}^2$. Consistency with the assumed one pion exchange nature of the 
leading singularity demands, however, that the $\pi N$ amplitudes are evaluated on-shell and that the pion 
virtuality only appears  through the pion propagator ({\it cf.} Fig.~\ref{BornScat}). 
\begin{figure}[h]
\centering
\includegraphics[scale=.35,clip]{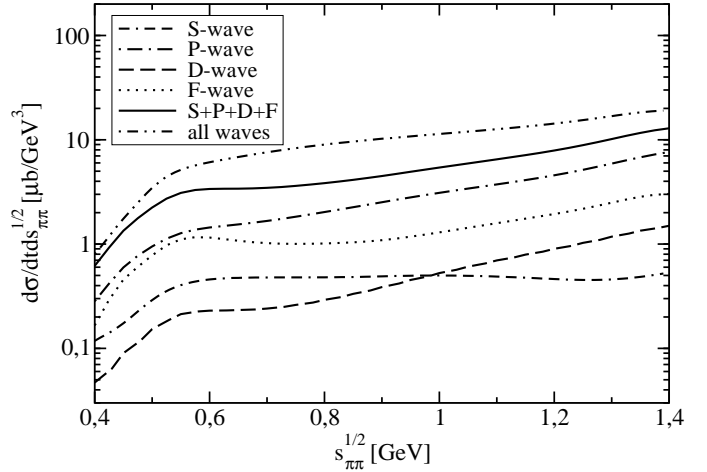}
\caption{Cross sections for low partial waves as compared to the cross section computed from the complete amplitude. The results are calculated without the final state interactions.}
\label{fig:dsdmdt-full}
\end{figure}
Even though the pion exchange is close to the physical region, because 
of the finite momentum transfer between the target and recoil nucleon,
 $t$ the  Deck amplitude gives a rather smooth function of $s_{\pi\pi}$. 
In Fig.~\ref{fig:dsdmdt-full} we compare individual cross sections 
computed for each of the four lowest partial waves ($S$,$P$,$D$,$F)$ 
of the Deck amplitude, with their incoherent sum in Eq.~(\ref{crs})  
and with the total, unprojected Deck amplitude (``all waves''). 
The calculation is done 
%shown in Fig.~\ref{fig:dsdmdt-full} 
 at photon 
energy $E_\gamma$ = 3.3 GeV and momentum transfer squared $t$ = $-0.55$ GeV$^2$.  
We observe that the convergence rate of the partial wave expansion is rather 
slow, so that the combined four lowest waves account for roughly 50$\%$ of 
the total contribution to the $s_{\pi\pi}$ intensity distribution. 
Moreover, the clear hierarchy of partial waves is visible, with  the odd 
partial waves being stronger than the even ones. This can be understood by 
considering the $\cos\theta$ 
and $\phi$ dependence of $T^+_{\lambda_1\lambda_2}$  and $T^-_{\lambda_1\lambda_2}$ in Eq.~(\ref{Deck}).  
  Changing  $\theta\to\pi-\theta$ and  $\phi\to\phi+\pi$ in the 
   second term of Eq.~(\ref{Deck})
and using Eq.~(\ref{charge-amp}) we see that the partial wave expansion in Eq. (\ref{pw-expansion}) can be rewritten as
\begin{IEEEeqnarray}{lCl}
%\nonumber
 {\cal M}_{\pi^+\pi^-}^{lm}=  && 
-e\int \!d\Omega \,{Y^\ast_{lm}}(\Omega)
\left(\frac{\epsilon_{\lambda}\cdot k_2}
{q\cdot k_2}-\frac{\epsilon_{\lambda}\cdot(p_1+p_2)}
{q\cdot (p_1+p_2)}\right) \nonumber \\
& & \times  \left[ T^{\frac{3}{2}}-\frac{(-1)^l}{3}(T^{\frac{3}{2}}+2 T^{\frac{1}{2}})\right].\label{pw-hierarchy}
\end{IEEEeqnarray}
It thus follows that in the case of even partial waves, $l=0,2,\dots$, 
  the dominant $\pi N$ isospin $3/2$ component is 
partially canceled while in odd ones it is enhanced, which explains 
 qualitatively the hierarchy observed in Fig.~\ref{fig:dsdmdt-full}.

{\it Numerical results.} 
%\section{Numerical results}
In general, in the kinematics discussed here, a minimal model for $\pi^+\pi^-$ photoproduction should contain two parts. One  corresponds to production of pion pairs from a spatially extended region and is given by 
  Eqs.~(\ref{eq_PRD-16}) and (\ref{eq_PRD-162}). We refer to this component as ``Deck+FSI''. The other, corresponds to production 
   from a spatially  compact source. For each partial wave the latter can be  parametrized by a short-range contribution given by, 
\begin{equation}
(A+B\,s_{\pi\pi})\,{\rm e}^{i\delta^I_l}\,{\rm sin}\delta^I_l
\label{corr}
\end{equation}
%where $\delta^I_l$ is the $\pi\pi$ phase shift used in
% Eq.~(\ref{eq_on-shell}) 
%with $I=0$ and 1 for the even and odd partial waves, respectively.
The term in the parentheses effectively parametrizes the smooth $s_{\pi\pi}$ dependence,  which in the physical region arises from exchanges of heavier mesons and/or quarks. This term is modified in the standard way by final state interactions in the $\pi\pi$ channel, where,  given the limited data range, we ignore inelastic effects. 
The free parameters $A$ and $B$ were fitted to experimental mass distributions extracted from the CLAS data. We compare predictions of the model with the mass distributions for low partial 
waves determined by the CLAS collaboration \cite{CLAS09}, which, to our knowledge, are the only available data on the di-pion partial-wave mass distributions.  
%
% Figure 3
%
\begin{figure}[h!]
\centering
\includegraphics[scale=.35,clip]{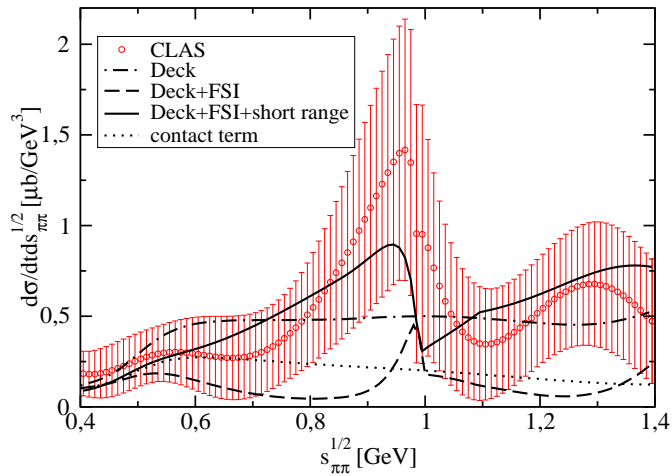}
\caption{$S-$wave double differential cross section at $E_\gamma$=3.3 GeV and 
$-t$=0.55~GeV$^2$. Dash-dotted line - pure Deck model; dashed line - Deck model 
with final state $\pi\pi$ interactions; solid line - Deck model with FSI and the short range term; dotted line - contribution of the contact term; 
 red points - CLAS fit to the experimental data. The error band shows the total uncertainty that combines the systematic and statistical uncertainties. (color online).}
\label{fig:dsdmdt-S}
\end{figure}

\begin{figure}[h!]
\centering
\includegraphics[scale=.35,clip]{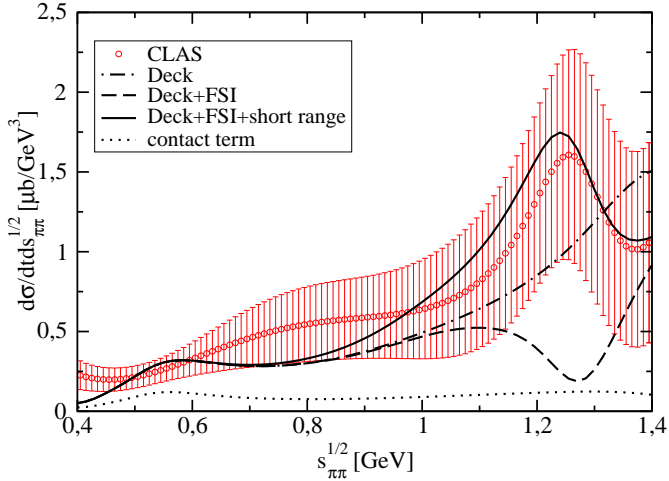}
\caption{$D-$wave double differential cross section at $E_\gamma$=3.3 GeV and $-t$=0.55 
GeV$^2$ with $M\leq$1. Dash-dotted line - pure Deck model; dashed line - Deck model 
with final state $\pi\pi$ interactions; solid line - Deck model with FSI and the short range term; dotted line - contribution of the contact term; 
red points - CLAS fit to the experimental data. The band shows the total uncertainty of the fit.  (color online).}
\label{fig:dsdmdt-D}
\end{figure}

In Fig.~\ref{fig:dsdmdt-S} we compare  model predictions with the 
experimental $S-$wave mass distribution which we denote here by CLAS fit as it was obtained from fitting the measured data~\cite{CLAS09}. It is clear that already the Deck amplitude alone gives the right magnitude of mass 
distribution and reproduces the mass dependence of background, {\it i.e} outside the 
 region of the $f_0(980)$ resonance. When the final state $\pi\pi$ 
interactions are taken into account (``Deck+FSI"), the resonant shape
 around 1 GeV,  is well reproduced. 
 Destructive interference between direct di-pion production and final state interaction {\it cf.} Eq.~(\ref{eq_PRD-16}) results in 
the mass distribution dipping below the experimental points in the whole energy region (see the discussion below Fig.~\ref{fig:dsdmdt-D} for more details). If, however, we include the short range component with parameters $A$=$-$14.5$\pm0.6$ GeV$^{-1}$ and $B$=2.7$\pm0.6$ GeV$^{-3}$ the fit fairly reproduces the mass distribution behavior both in resonance region and outside.  
The slightly different invariant mass behavior of our predictions above 1 GeV in comparison  with the CLAS fit can be attributed to the absence of the $K\bar{K}$ channel in the model. Another point we would like to discuss here is a contribution of the correction term 
in Deck amplitude, Eq.~(\ref{Deck}) required for gauge invariance, 
 typically referred to as a contact term (even though in our case 
 it is not local). 
In Fig.~\ref{fig:dsdmdt-S} we show the contribution of the 
contact term in Eq.~(\ref{Deck}) (the dotted line). It is apparent that in the 
region around 0.6 GeV this contribution reveals a small  enhancement in the mass 
distribution. This enhancement is also seen in the curve obtained from the ``Deck+FSI" amplitude. One can also say that the contribution is relatively large in the $S$ wave.
\begin{figure}[ht]
\centering
\includegraphics[scale=.35,clip]{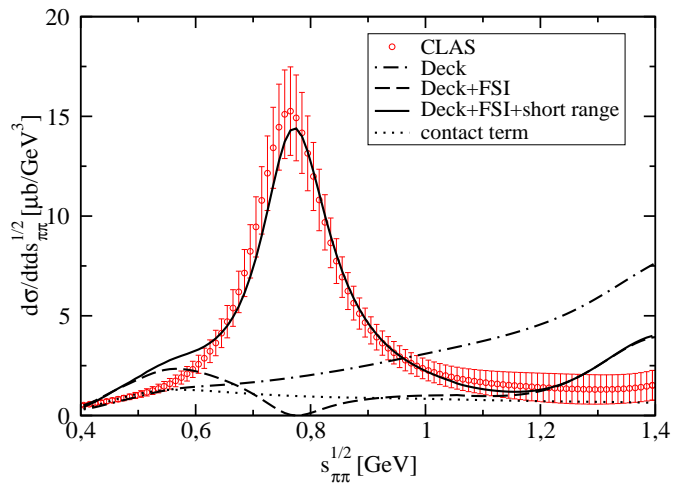}
\caption{$P-$wave double differential cross section at $E_\gamma$=3.3 GeV and $-t$=0.55 
GeV$^2$. Dash-dotted line - pure Deck model; dashed line - Deck model 
with final state $\pi\pi$ interactions; solid line - Deck model with FSI and the short range term; dotted line - contribution of the contact term; 
red points - CLAS fit to the experimental data. The band shows the total uncertainty of the fit.  (color online).}
\label{fig:dsdmdt-P}
\end{figure}

In Fig.\ref{fig:dsdmdt-P} we show the CLAS $P-$wave mass distribution compared to our model predictions. The overall agreement of data with the full model (Deck+FSI+ short range), especially in the resonance region, is good. 
However, the long-range component with final state interactions (Deck+FSI) produces a minimum rather than the maximum at the resonance energy.
%However, when only the long range (Deck) component is taken into account the amplitude develops the minimum rather than maximum at the resonance. 
Thus the peak of the $\rho(770)$ resonance, as expected is due to the short range production. Specifically we find $A$=48.9$\pm$1.6 GeV$^{-1}$ and $B$=-24.3$\pm$2.0 GeV$^{-3}$. A comparison 
of the fitted values of the $A$ and $B$ parameters for the $S$ and $P$ waves implies that the relative contribution of the short range component of the amplitude is much larger in the $P$ wave, as expected for the standard $q\bar{q}$ state. 
Small deviations from the data can be observed in the near threshold region and for masses well above the $\rho(770)$ mass. The near threshold discrepancy results from a small enhancement in the contact term magnified by final state interactions.
An alternative model for the $P-$wave photoproduction of $K\bar{K}$, based on the pomeron exchange dominance, can be found in \cite{Lesniak}, which also applies to the $\pi^+\pi^-$ case.
%
% Figure 5
%

In Fig. \ref{fig:dsdmdt-D} we show our model results  compared to CLAS $D-$wave 
mass distribution. It is important to note that following the experimental analysis 
we take into account only the amplitudes where the magnetic quantum number $M$ of 
the $\pi\pi$ system (equivalent to the helicity in the chosen frame of reference) 
is smaller than 2. Similarly as in the $S$ wave, the model gives the right magnitude of the experimental points even for the pure Deck amplitude.  
Recall that this result is parameter free, contrary to the results in Ref.~\cite{PRD-13} that were fitted to the experiment.
 Inclusion of the final state 
interactions, similarly as in the $P-$wave, results in developing the minimum rather than the maximum for the  invariant masses around the $f_2(1270)$. 
This different pattern in the $S$ and $D$ waves can be understood from behavior  of the isoscalar $\pi\pi$ phase shifts~\cite{BKN2016}. 
 The production amplitude in Eq.~(\ref{eq_PRD-16}) is dominated by 
the term proportional to $\cos\delta^0_l$, which comes from the square brackets in Eq.~(\ref{eq_PRD-16}).  Then the minimum in the $D$ wave is due to the $\pi\pi$ phase shift passing $\pi/2$ at about 1.25 GeV.
In the $S$ wave the phase first passes $\pi/2$ at about 
0.85 GeV as seen in Fig.~\ref{fig:dsdmdt-S}  
for ``Deck+FSI''. When the $S$ wave phase shift reaches  $\pi$  at  $\sqrt{s_{\pi\pi}} \sim 0.95 \mbox{ GeV}$ it produces a maximum. 
The model agrees much better with the $D$-wave data if we include the 
short range component with parameters $A$=$-$24$\pm11$ GeV$^{-1}$ and $B$=10$\pm7$ GeV$^{-3}$. It is obvious from Eq. (\ref{corr}) that the $D$ wave resonates at $\sqrt{s_{\pi\pi}} \sim 1.25 \mbox{ GeV}$ (so, the overall amplitude behavior is quite analogous as in the $P$-wave). 
The contribution of the contact term is not so  important in the $D$ wave as in the $S$ wave 
but it also reveals a tiny bump below 0.6 GeV that is apparent in the full result 
(the solid line).
\begin{figure}[t]
\centering
\includegraphics[scale=.35,clip]{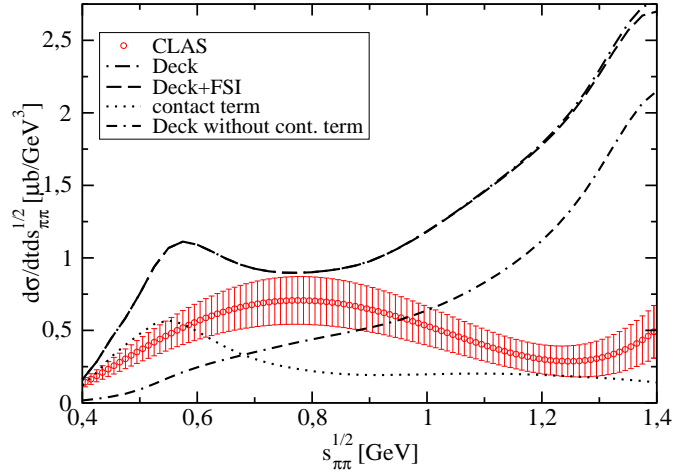}
\caption{$F-$wave double differential cross section at $E_\gamma$=3.3 GeV and $-t$=0.55 
GeV$^2$ and $M\leq$1. Dash-dotted line - pure Deck model; dashed line - Deck model 
with final state $\pi\pi$ interactions; dotted line - contribution of the contact 
term; double-dash-dotted line - Deck without the contact term; red points - CLAS fit to the experimental data. The band shows the total uncertainty of the fit.  (color online).}
\label{fig:dsdmdt-F}
\end{figure}

In Fig. \ref{fig:dsdmdt-F} we compare the model prediction with the $F-$wave mass 
distribution measured by CLAS. A discrepancy is observed throughout the entire  mass region. Moreover, the effect of the final state interactions in the $F$ wave is 
negligible, which results from  very small values
 of $\pi\pi$ partial waves. On the other hand the effect of the contact term is relatively large here and it explains the bump around 0.6 GeV. It is apparent that a form of the contact term 
is responsible for the excess in the mass distribution below 0.8 GeV, 
as indicated by the double-dash-dotted line. As the contribution of the contact term is flat 
it cannot contribute to the rising distribution at high masses. 
 
%\section{Conclusions and outlook}
{\it Conclusions and outlook}. 
With the model discussed in this paper we have calculated mass distributions for 
various partial waves in photoproduction of the $\pi^+\pi^-$ pairs on 
the proton. In our approach we combine the Deck model, which accounts for the extended source mode of the photoproduction, with the SAID parametrization 
of $\pi N$ scattering amplitudes. This part of the model is essentially parameter free. 
Thus, we have probed the dominant exchange mechanism of the reaction at forward angles that 
is given by the one pion exchange in the $t_{\gamma \pi}$ channel. We also took into account the compact
source mode of the reaction, which based on the general grounds can be parametrized by a smooth function. In this respect we have used a first order polynomial in $s_{\pi\pi}$. When we include the final state $\pi\pi$ interactions in the model, we obtain 
the $\pi\pi$ mass distributions which for low partial waves are in good 
agreement with CLAS measurements made at $E_\gamma$=3.3 GeV. 
Predictions of the model agree well with the experimental fact that the $S$ and $D$ waves are 
dominated by isoscalar $f_0(980)$ and $f_2(1270)$ resonances, respectively, whereas the $P$ wave is dominated by the isovector $\rho(770)$ resonance. Moreover, we observe
that the compact source component of the resonant amplitude in $P$ and $D$ waves is larger than this same component for the $S$ wave (compare eg. the values of the corresponding $A$ and $B$ parameters). 
This is in line with the expectation that while the $\rho(770)$ and $f_2(1270)$ are typical $q\bar{q}$ resonances, the $f_0(980)$ is rather more loosely bound system of four quarks.
In the $F$ wave we observed the discrepancy between 
CLAS measurements and model predictions. At small invariant masses  we attribute  this 
discrepancy to a specific form of the contact term adopted from~\cite{Pumplin}. We observe 
a general hierarchy of the partial waves resulting from the pure Deck model, namely that 
the even partial 
waves are weaker than the odd ones which can be qualitatively inferred from Eq. (\ref{pw-hierarchy}). 

A similar analysis using the Deck (Drell) mechanism driven by the kaon exchange 
for the $K\bar{K}$ photoproduction was performed in \cite{Sibirtsev}. 
In their analysis the authors took into account the full $KN$ and $\bar{K}N$ 
scattering amplitudes showing that the kaon exchange mechanism alone is 
not sufficient to describe the data on the $K^+p$ and $K^-p$ invariant 
mass spectra. 
The reaction mechanism was therefore extended by adding the $K^*$ exchange 
with a large coupling to the $\Lambda(1520)$ resonance and a better description of   the invariant mass spectra was achieved. 
Our findings are consistent in that the  
reaction mechanism based only on the long range 
mode is not enough to get a realistic description of  the data. The two-pion photoproduction on the nucleon
was also studied at small energies ($E_\gamma < 1.5$ GeV) 
 in ~\cite{Fix} based on an effective Lagrangian approach. 
To achieve a satisfactory description of the data on total cross sections 
the authors included many baryon resonances in the $s$-channel with the mass below 1.8 GeV. In the 
$t$-channel, exchanges of heavier mesons ($\sigma$ and $\rho$) were included showing that also in this approach far-away singularities do play 
important role.

Our formalism allows for systematic refinements of the model. These include the coupled channel effects (which we 
expect to be important especially for the isoscalar $S$ wave), off-shell effects and inclusion of other $t$-channel 
exchanges. In order to use the model in the full kinematic region accessible for GlueX and CLAS12 energies, the SAID 
$\pi p$ amplitudes must be supplemented with amplitudes applicable for $\pi p$ CM energies beyond 2 GeV.

\section*{Acknowledgments}
Authors acknowledge fruitful discussions with members of the JPAC collaboration.
This work was supported by the Grant Agency
of the Czech Republic under the grant No. P203/15/04301,  the U.S.~Department of Energy under grants No. DE-AC05-06OR23177 and No. DE-FG02-87ER40365, and the U.S.~National Science Foundation under award numbers PHY-1507572, PHY-1415459 and PHY-1205019.

\end{document}